\documentclass[prE,twocolumn]{revtex4}
\usepackage[T1]{fontenc}
\usepackage[cp1250]{inputenc}
\usepackage{color}
\usepackage{graphicx}
\usepackage{hyperref}
\begin{document}

\title{Hate networks revisited: time and user interface dependence study of user emotions in political forum}

\author{Pawel Sobkowicz}
\affiliation{KEN 94/140, Warsaw, Poland}
\email{pawelsobko@gmail.com}
\homepage{http://countryofblindfolded.blogspot.com}

\author{Antoni Sobkowicz}
\affiliation{Mechatronics Department,
Warsaw University of Technology
Sw. Andrzeja Boboli 8,
02-525 Warsaw, Poland}

\begin{abstract}
The paper presents analysis of time evolution within am Internet political forum, characterized 
by large political differences and high levels of emotions. The study compares samples of 
discussions gathered at three periods separated by important events.  
We focus on statistical aspects related to emotional content of communication and  
changes brought by technologies that increase or decrease the direct one-to-one discussions.
We discuss implications of user interface aspects on promoting
communication across a political divide.
\end{abstract}

\maketitle

\section{Introduction}

Internet discussion fora are very fertile grounds for research of human 
communication patterns and social 
structures \emph{in statu nascendi} -- that is when the links between communicating people can 
literally be observed as they form, together with the content, timing structure and 
emotional tone of the messages. Many Web sites provide tools that allow the users to  express 
their views, comment important topics and reply to posts by other users.
Resulting  discussions are often only slightly moderated, allowing various kinds of expressions, 
ranging from elaborate texts to single words or emoticons; 
from polite discussion to exchanges of obscenities.  
The Internet allows the users to remain relatively anonymous, thus they are  free 
from anxiety of expressing extreme views due to possible 
retributions typical face to face contacts or formal correspondence. 
At the same time, certain stability of 
nicknames allows recognizability within the discussion platform, 
so that social networks may 
form, grow and evolve. 
The resulting social networks stretch across geographic distances, 
social status, age and political divides.
The discussion fora have attracted significant research attention:  
\citet{mullen06-1,kelly06-1,kelly09-1,schuth07-1,schuth07-2,
wu08-1,wu08-2,gomez08-1,grabowski08-1,grabowski09-1,kulakowski09-1,
tsagkias10-1,yunjunglee10-1,schweitzer10-1,si10-1,si10-2,
ding10-1,ding10-3,chmiel11-1}. These works described multiple characteristics of user behavior, both from statistical point of view, describing social network properties, and from social dynamics perspective (opinion spreading, emotions expressed by the users). 
There are also works focused on discussions spurred by personal blogs 
(\citet{jeong03-1,jeong05-1,mishne06-1,leskovic07-1}) and of networks formed by the blogs 
themselves (\citet{adamic05-1,trammell06-1,hargittai07-1}).

The importance of such studies results not only from the freedom of expression mentioned above, 
but also from the variety of motivations driving specific discussion fora. These may vary from 
helpful assistance (e.g. in computer technology), virtual gatherings of aficionados of particular 
activity (sports, music fans, entertainment\ldots), reviews and opinions concerning specific 
products or services (hotels, gadgets, books\ldots) to political discussions.

In our previous work  (hereafter referred to as \textbf{Paper~I}, \citet{sobkowicz10-1}), 
we have presented results of studies of Internet discussions powered by strong negative feelings, 
within highly polarized Polish political environment. 
Paper I compared statistical properties of such interactions with those of less contentious topics 
(for example sport or computer technology discussions) and presented a simple simulation model, 
in which large role was given to pairwise exchanges of comments between individual participants. 
Our motivation was to see, if there are particular properties of social networks 
that are formed by linking representatives of conflicted sides. Such hate based networking 
is quite unusual outside the Internet, because in 
real life voluntary social links are mostly based on common interests and views. 
We have found that the network based on negative emotions may be quite extended, 
and that statistical behavior
shows many similarities to networks based on cooperation and shared interests, 
e.g. power law distribution of indegree and outdegree. These observations are in 
agreement with those of Chau et al. (\citet{chau06-1,chau06-2}) who studied the network 
structure of `hate groups' and Chmiel at al. (\citet{chmiel11-1}), who analyzed a 
large dataset of discussions of the BBC political, religion and news fora.

The goal of this work is to extend the scope of Paper I in two directions. 
The first is to broaden the scope: we monitor the same news site for over two years, 
looking for elements that remain stable and those that change. The second direction 
is to study effects of change in visual presentation of the discussions and other 
elements of the user interface on resulting social networks, emotion levels and capacity to communicate.

During the two years that have passed since gathering of data for Paper I, 
the political split in Poland has significantly increased. Tragic crash of the plane 
carrying the President Lech Kaczynski resulted in snap presidential elections. 
As no candidate received a majority of votes in the first round, a second round was held 
on 4 July 2010 in which Bronislaw Komorowski, candidate of Platforma Obywatelska 
(Civic Platform, PO) defeated Jaroslaw Kaczynski, candidate of Prawo i Sprawiedliwosc 
(Law and Justice, PiS), twin brother of the  president who died in the catastrophe. 
The elections had relatively high turnout (55.31\%) and have split the voters almost 
half in half (53.01\% to 46.99\%).\footnote{For detailed data see Web page of the 
National Electoral Commission 
(\url{http://prezydent2010.pkw.gov.pl/PZT/EN/WYN/W/index.htm}).}

High emotional content of pre- and post-election discussions resulted in apparently 
unbridgeable split of the society at all levels. Voting data of the National Electoral
 Commission show that there is significant geographical correlation between the voter 
 preferences and city size and industrialization level. In many cases direct contact 
 between the supporters of the two camps is limited. On the other hand the Internet, 
 allows such contact, with little or no  limitations on content or attitude. 

In the current paper we shall denote the four sets of comments by their  time of origin: 
January 200  (JAN09, covered in Paper I), second half of July 2010 (JUL10) 
and two datasets gathered in February 2011 (FEB11 and FEB11Q). 
The difference between the two latter datasets shall be explained in next section. 
Our goal is to examine 
statistical aspects of the discussions not covered in Paper I (such as emotion distribution) 
but also to look into general changes brought by the passage of time. In this aspect, the JUL10
dataset stands out, as it has been gathered just after the loss of the elections by the PiS candidate, 
when his supporters were in highly emotional state, some of them denying legitimacy of the 
voting result, some accusing the president-elect of treason. 

\section{New data description}

\subsection{Data sources}
As in Paper I, we have gathered our data from discussion fora related to news items published in 
category Politics by the Internet branch of the largest Polish newspaper, \url{gazeta.pl}. 
While the news source  is not neutral (the newspaper has clearly an anti-PiS stance), 
the discussion participants from both camps use openness of the forum as a convenient `battleground'. 
Posting comments requires a registered user account, and while other users see only the 
other users' nicknames, the privacy policy warns users that IP address and other data are 
logged and may be given to the relevant state agencies in the case of lawbreaking.
There is possibility for a single person to register multiple times, but as we have no access to
such information we treat each nicknames as a separate user.

In addition to posting comments, registered readers may score the other comments via simple 
thumbs up/thumbs down mechanism (such evaluations are part of the JUL10 and FEB11 datasets). 
The current score is displayed along the post, in boldface, green/red font to increase visibility 
(see Fig. \ref{fig:NewUserInterface}). 

Non-neutrality of the forum, resulting from the political sympathies of the newspaper 
that provides the basic news items, is strengthened by automatic hiding of posts that have 
strong negative score. There is only `comment hidden' and score visible during normal browsing 
and the post may be made visible by individual clicking on dedicated link.  
It should be noted that this visual hiding of the comment is separate  action from administrative 
deleting comments reported as illegal, aggressive etc. The latter option is exercised 
relatively infrequently (for less than 1\% of the posts). In our analysis we have found that 
the automatically hidden comments are not significantly more abusive than the rest, the 
distinction is that they represent minority view. As a result, the first impression that 
viewers of the web page have is of greater uniformity of political opinions, due to combination of 
administrative mechanisms and particular mix of participants.

\subsection{Effects of user interface changes on communication characteristics}
In addition to searching for effects of passage of time, we have identified another 
significant  significant factor that might have influenced the user behavior. 
At the time of writing of Paper I, the forum interface allowed users to post replies to 
other users' comments with a single click. This resulted in the observed high proportion 
of exchanges of posts between pairs of users. The graphical form of presentation of the 
discussion threads favored easy recognition of such exchanges by other users, which 
increased the ratio of comments directed to comments (rather than to the original news stories)

\begin{figure*}[ht]
\includegraphics[height=150mm]{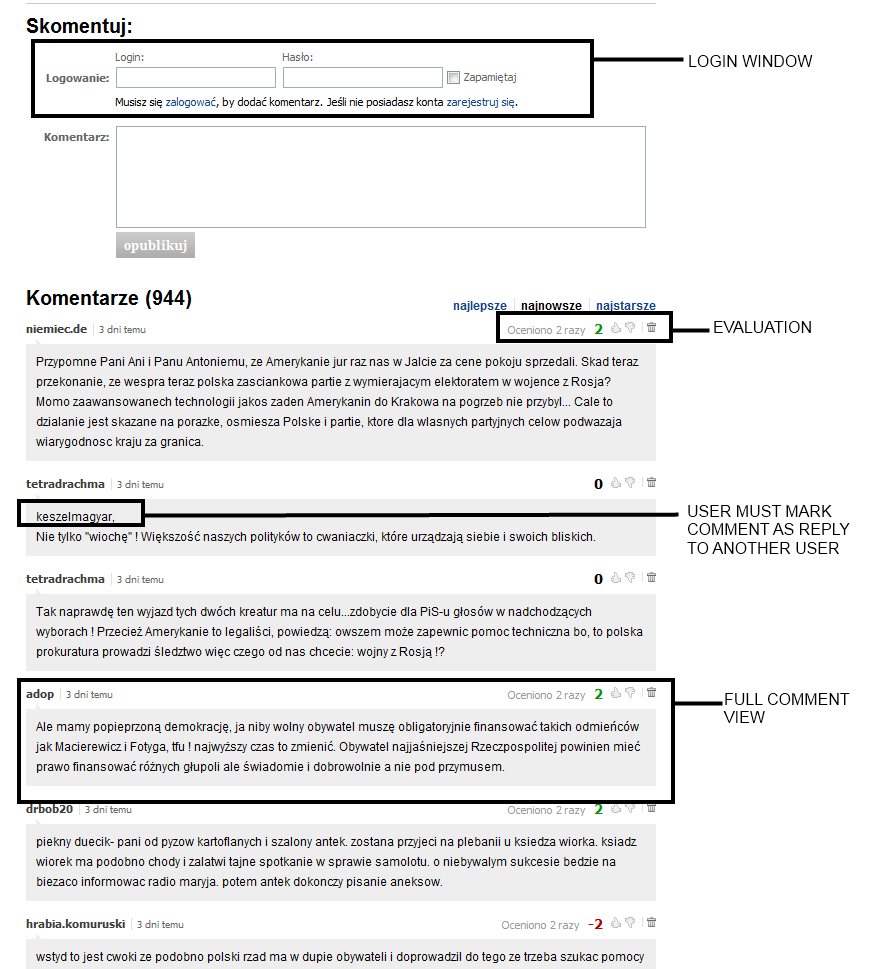}	
	\caption{`New' user interface to news discussion forum, \url{www.gazeta.pl}. The interface shows each comment in full and allows one-click evaluation (thumbs-up, thumbs-down), together with the status of previous evaluations. To reply to a specific comment written by someone else, users resort to direct mentioning of the author in the text of the comment. No tree structure is visible. The discussion forum is placed directly beneath the full news item. }
	\label{fig:NewUserInterface}
\end{figure*}

 Since mid 2010, the forum has technically split into two branches. The first, \url{www.gazeta.pl} (Fig.~\ref{fig:NewUserInterface}), does not allow direct individual responses to specific posts. Thus, there is no longer a simple mechanism supporting quarrels that we have described. The change has been criticized by many users, some of them calling for boycott of the forum. The critics openly state that they want to interact with the other users. Interestingly, these criticisms come from users from both political camps, and are probably the sole topic on which the two groups agree. Throughout the paper we shall refer to this as the `new interface'.

\begin{figure*}[ht]
\includegraphics[height=150mm]{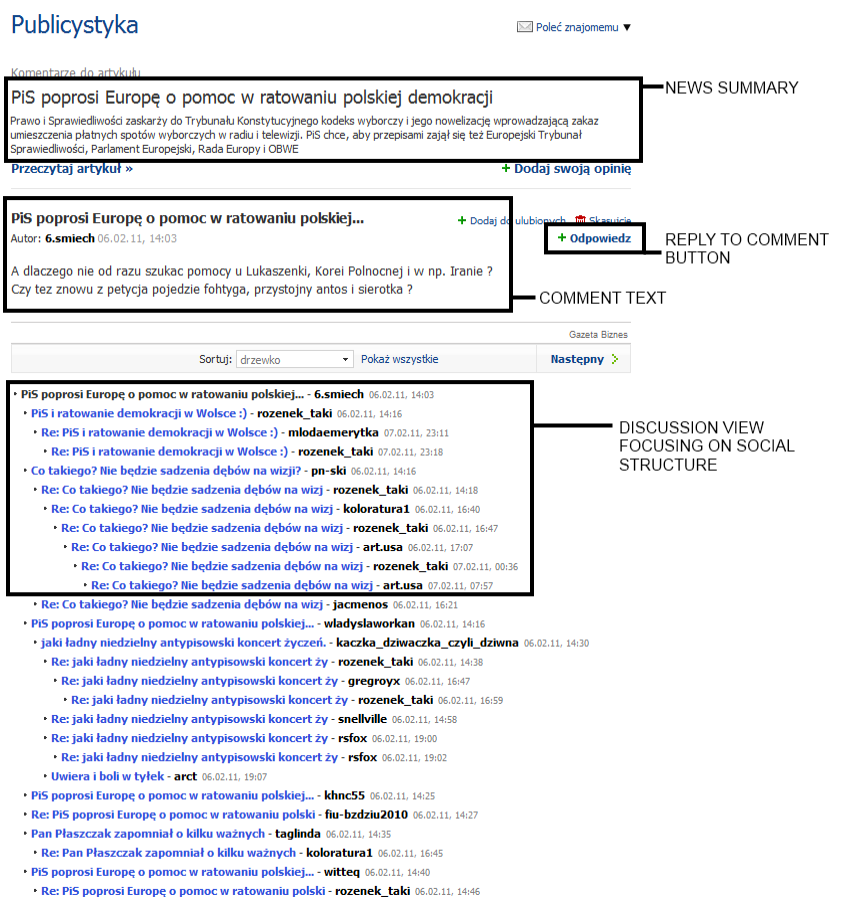}	
	\caption{`Old' user interface to news discussion forum, \url{forum.gazeta.pl}. The interface facilitates replies to individual comments and visualize the tree structure of the discussions. The screenshot presented here highlights and example of a quarrel between two users. On the other hand to read each comment in full the viewer must click on particular link. Discussion forum is placed beneath a summary of the news item, rather than the original story, where only number of comments is provided.}
	\label{fig:OldUserInterface}
\end{figure*}

The second, less popular site, \url{forum.gazeta.pl} (Fig.~\ref{fig:OldUserInterface}), has preserved the 
old capacity of one-click replies (and the quarrels that we have observed in Paper I). 
Tree-like structure of the posts is prominently visualized (rather than flat time-dependent
sequence), so that the most active exchanges between pairs of commentators 
are immediately visible and draw a lot of attention from other users.
We shall refer to it as the `old interface'.
It is worth noting that the base news stories are essentially the same on both datasets. 
This situation gives us unique opportunity to study the influence of presentational 
aspects in Internet communications on the user behavior and expression of emotions.
To focus on differences due to the user interface we gathered, in February 2011, two parallel datasets 
corresponding to  new interface: \url{www.gazeta.pl} -- dataset FEB11 and to the old interface:
\url{forum.gazeta.pl} -- dataset FEB11Q.

Despite the lack of tools facilitating  exchanges of posts between users in the new forum, 
we observed a variety of impromptu editing tricks aimed to indicate that comments are related 
not to the main story but to other users. Typically, such posts are often started with reference 
to someone's nickname (often preceded by @ sign for visibility). Sometimes the reference is 
hidden inside the comment. In the analyzed sample we had to identify such discussions 
through human reading. 
The process could not be automated, as users frequently misspelled other users nicknames. 
Each of the samples analyzed consisted of several thousand posts in a few tens of threads. 
To analyze the social networking statistics we have focused on discussions with more than 50 posts, 
typically in the range of 100--500. 
Selection of the threads used for analysis was based solely on their size, with no 
pre-screening of content of comments. We note here that the size of a 
discussion obviously depends on the `hotness' 
of the commented topic. 
Some news stories were in themselves quite provocative (e.g. commenting negatively on 
prominent PiS politicians) so that one could expect that they would rise a lot of comments. 
But sometimes long threads resulted from relatively low profile news.
The use of only selected, long discussions in our analysis 
makes direct comparisons of post statistics with data gathered from \textbf{total} 
records of other Internet discussion fora (\citet{si10-2,si10-1,ding10-1,ding10-3,chmiel11-1}) 
impossible. For example, our datasets would have smaller number of posts with no links to other posts 
simply due to the fact that we avoided news items which generated very small discussions.

In the discussions using the new user interface we were able to identify only a 
few extended quarrels involving 
pairs of users, the longest comprising of 4 posts. This is drastically shorter than the exchanges 
in JAN09 and FEB11Q data, where we observed many exchanges longer than 10 consecutive posts. 
This is  obviously due to technical properties of the new portal which make 
such exchanges more difficult to maintain -- 
a user has to watch for replies to his/her own comment without graphical guideline to help,
which makes responding more difficult. 
The lack of visibility of quarrels also diminishes interest in joining-in by onlookers. 
As a result, the new interface promotes many more self-contained comments, 
which do not relate to other users.

\section{Network analysis}

\subsection{User and comment network properties}
Within each thread, each discussion forum, we may treat users as forming a directed social 
network, with links provided by comments directed at other users. Comments attached to source news 
(and not to other users' comments) may be considered as marking the presence of active but isolated nodes.

\begin{table*}[ht]
\begin{tabular}{|p{6cm}|p{2cm}|p{2cm}|p{2cm}|p{2cm}|}
\hline \rule[-2ex]{0pt}{5.5ex}   & JAN09 & JUL10 & FEB11 & FEB11Q\\  \hline
\hline \rule[-2ex]{0pt}{5.5ex} Forum interface & OLD & NEW & NEW &  OLD \\ \hline
\hline \rule[-2ex]{0pt}{5.5ex} Number of threads & 47 & 27 & 27 &  50\\ 
\hline \rule[-2ex]{0pt}{5.5ex} Number of posts & 7592 & 7179 & 6447  &  4591\\ 
\hline \rule[-2ex]{0pt}{5.5ex} Number of users & 1613 & 2752 & 2187 & 1527\\ 
\hline \rule[-2ex]{0pt}{5.5ex} Number of links between users (percentage of posts) & 4754 (62.6\%) & 770 (10.7\%) & 1172 (18.2\%) & 2286 (49.8\%)\\ 
\hline \rule[-2ex]{0pt}{5.5ex} Largest connected component (percentage of the users)& 1106 (68.5\%) & 440 (16.0\%)& 589 (26.7\%)& 947 (62.0\%)\\ 
\hline \rule[-2ex]{0pt}{5.5ex} Isolated users (percentage of the users)& 457 (28.3\%) & 2156 (78.3\%) & 1507 (68.9\%) & 522 (34.2\%) \\ 
\hline \rule[-2ex]{0pt}{5.5ex} Multi-edge pairs (percentage of the users)& 718 (44.5\%)& 92 (3.3\%) & 160 (7.3\%) & 310 (20.3\%)\\ 
\hline 
\end{tabular}
 \caption{Comparison between JAN09, JUL10, FEB11 and FEB11Q network properties. 
 JAN09 and FEB11Q facilitate user-to-user exchanges, and show much greater proportion of links 
 and pairs of users connected by multiple links. The difference is especially visible in the number of 
 isolated users (i.e. users who posted a comment directed at the main news story, without 
 commenting other users) and in the relative size of the largest connected component of the user network, 
 which reaches over 60\% in the old, quarrel-promoting, interface. }
	\label{tab:JAN09JUL10compareNetwork}
\end{table*}
{Table~\ref{tab:JAN09JUL10compareNetwork} shows, for each dataset, the basic network parameters.} 
The old interface with its one-click reply mechanism facilitated extended network formation.
For both JAN09 and FEB11Q datasets the largest connected component comprised of more than
60\% of the users. One can think in these cases about forming a percolation network for
information travel among the users. And we recall here that the users are largely coming from opposing political camps. Thus the existence of such network shows that conflicted users at least see the arguments and narratives used by their opponents. On the contrary, much weaker network connection of the new forum with majority of users posting comments which do not relate to other users might indicate their focus on their own viewpoint only.

The new user interface has diminished the possibility of user-to-user communication. 
The majority of comments are now directed at news source. Many users are thus unconnected to others. 
With a similar number of posts, the largest connected component is almost three times smaller, 
there are more than 4 times more isolated users and 6 times less links. 
In Paper I we have identified pairwise exchanges of comments as the driving mechanism of 
network formation. This is well represented by a large number of pairs of users linked by 
multiple connections (multi-edge pairs). Between JAN09 and JUL10 the 
number of such pairs has fallen 
8 times. 
The FEB11Q data preserve some of the highly networked characteristics of the JAN09 dataset, 
e.g. significant size of the largest connected component, small percentage of isolated users.
We attribute smaller number of links to the 
lesser popularity of the forum, hidden, as we noted, deeply within the newspaper Web site.

Overall, our datasets contained 6404 users, defined as distinct nicknames. 
Out of these, 5132 were present in only 
one of the sets, 940 in two of them, 274 in three and 58 users participated in 
discussions contained in all four datasets. Presence in more than one set of data, 
which means long term presence in the forum, is correlated with general activity. 
The average number of posts per user for the small core of 58 users
was 27.2, compared to the overall average of 4.03.  For the users who were found in at least three
datasets, the average number of posts was 16.6.
Another way of looking at the extended activity measures is to check the most active users (as given by 
the overall number of posts). From the top ten, characterized by an average of 152 posts per user,  
two were present in only one set, two in  two datasets, four in three and two in all four.

\begin{figure*}[ht]
\includegraphics[height=160mm]{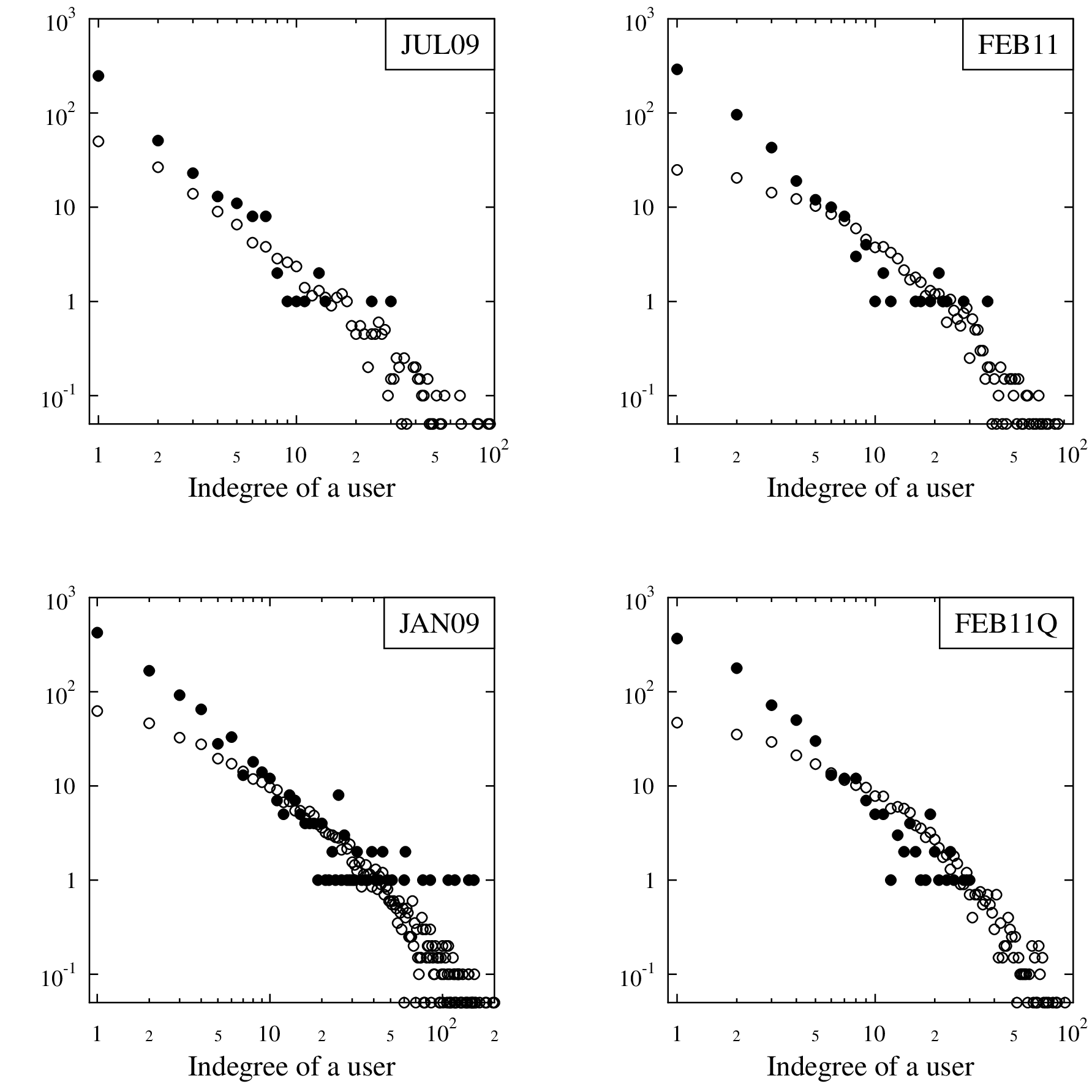}	
	\caption{Comparison of indegree  of users in various datasets (clockwise from bottom left: JAN09, JULY10, FEB11 and FEB11Q). Filled circles are normally binned observation data. Open circles are averages from 20 runs of simulations.} 
	\label{fig:indegree4}
\end{figure*}

\begin{figure*}[ht]
\includegraphics[height=160mm]{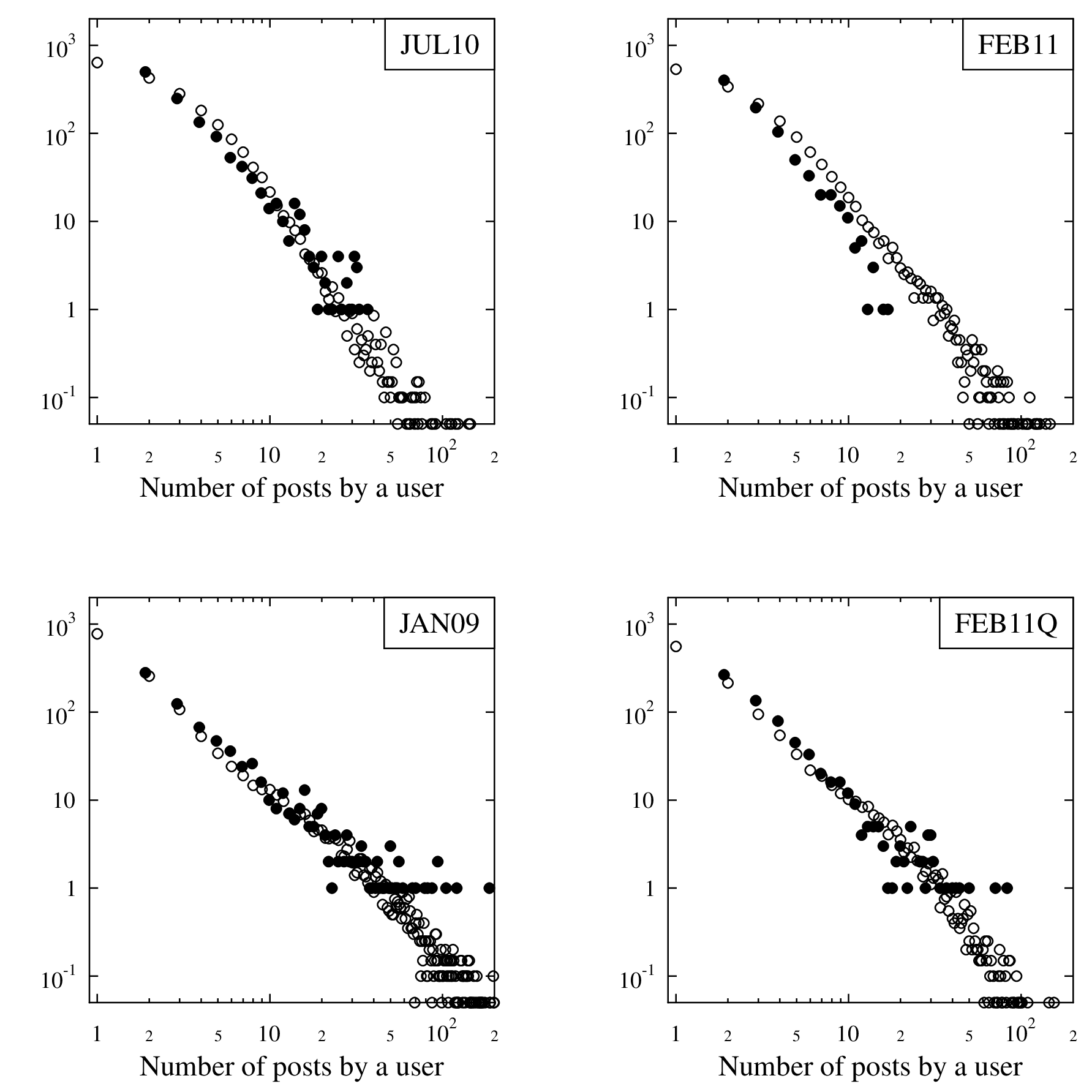}	
	\caption{User activity measured by number of posts written by the user (clockwise from bottom left: JAN09, JULY10, FEB11 and FEB11Q). Filled circles are normally binned observation data. Open circles are averages from 20 runs of simulations.}
	\label{fig:userposts}
\end{figure*}

\begin{figure*}[ht]
\includegraphics[height=160mm]{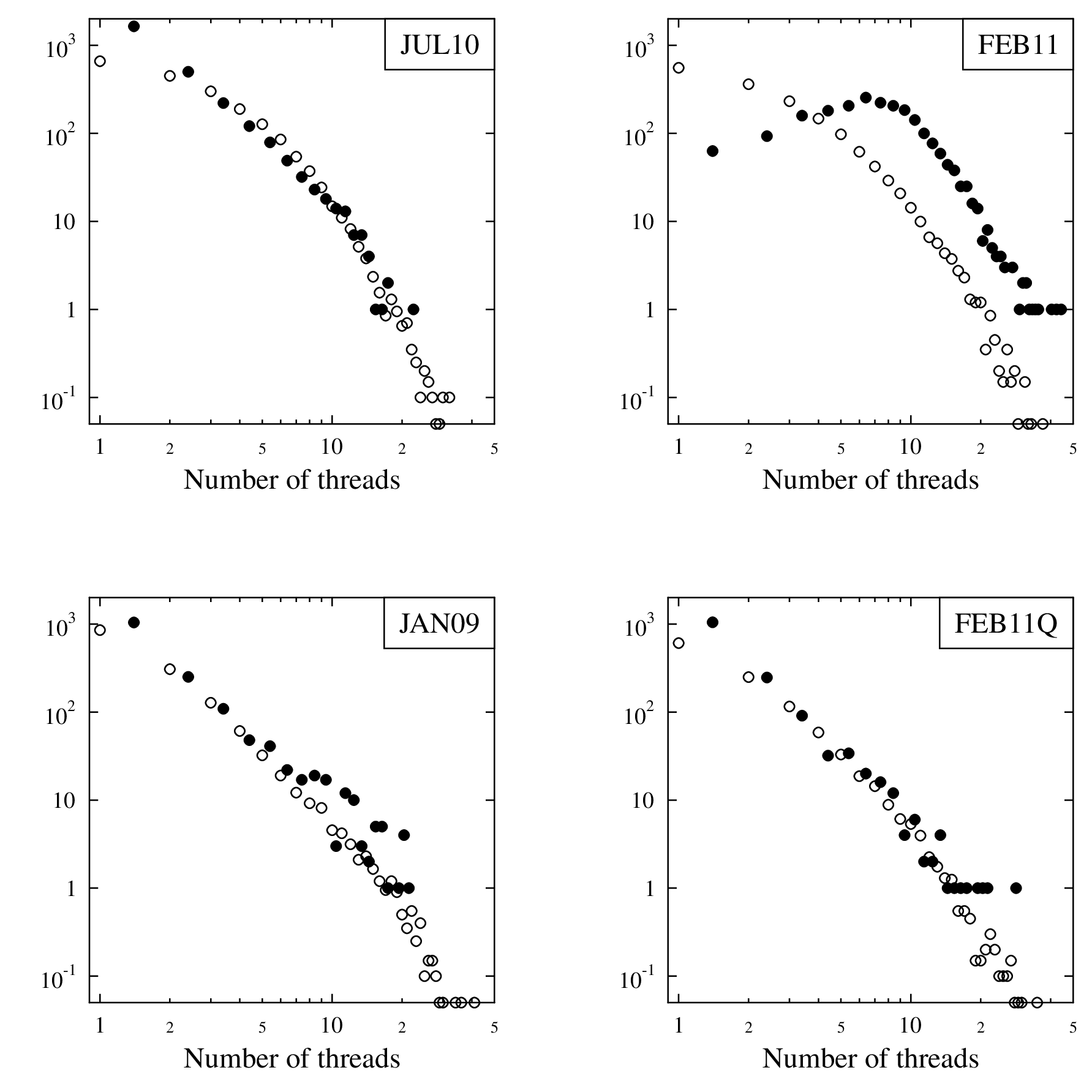}	
	\caption{User activity measured by number threads that users participate in  (clockwise from bottom left: JAN09, JULY10, FEB11 and FEB11Q). Filled circles are normally binned observation data. Open circles are averages from 20 runs of simulations. The origin of the deviation of FEB11 data from power law is unknown.}
	\label{fig:userthreads}
\end{figure*}

Figures \ref{fig:indegree4}--\ref{fig:userthreads} 
present distributions characteristic network measures of user activity for the four datasets. 
We focused on user indegree, number of posts written by a user and number of discussions 
he/she participated in. Most of these distributions are relatively well described by power 
laws, the only significant deviation is in FEB11 data on number of threads, which shows 
unusual behavior at low range.

The number of posts written by users and network outdegree should not be confused: 
as noted in Table \ref{tab:JAN09JUL10compareNetwork} many of the comments  do \textbf{not} 
connect two users, being addressed to the original news article, so they are not counted in 
the usual network analysis. For the new interface user-to-user comments are only a small 
part of the total number of posts.

\subsection{User political affiliation statistics}

Statistical properties of the comment fora may be analyzed from two points of view: looking at users and at discussion  threads.
 
The first element is the relative number of  participants from the opposing camps. 
As we have already noted, the news source (\url{www.gazeta.pl}) could hardly be called neutral. 
It shows strong pro-PO sympathies,
actively participating in election campaign. 
Not surprisingly, supporters of PO form the majority of readers and commentators. 
Table \ref{tab:Affiliations} summarizes the ratios of posts and participating users identified
as supporters of the two major combating parties and those with unknown sympathies.  
In all cases the PO supporters formed roughly 50\% or more of the participating users, PiS being 
a significant minority. This polarization of the forum participation has been observed by the 
users themselves,  and often mentioned in the posts.

%\begin{figure*}[ht]
%\includegraphics[height=100mm]{JULY2010_threads_lenght}	
%	\caption{Distribution of lengths of threads chosen for analysis in JUL10 dataset. Right panel presents histogram of the thread length in the threads selected for analysis.}
%	\label{fig:threadlength}
%\end{figure*}

\begin{table*}[ht]
\begin{tabular}{|p{6cm}|p{2cm}|p{2cm}|p{2cm}|p{2cm}|}
\hline \rule[-2ex]{0pt}{5.5ex}   & JAN09 & JUL10 & FEB11 & FEB11Q\\  \hline
\hline \rule[-2ex]{0pt}{5.5ex} Forum interface & OLD & NEW & NEW &  OLD \\ \hline
\hline \rule[-2ex]{0pt}{5.5ex} PO users & 62\% & 78\% & 62\% &  46\%\\ 
\hline \rule[-2ex]{0pt}{5.5ex} PiS users& 22\% & 15\% & 20\%  &  29\% \\ 
\hline \rule[-2ex]{0pt}{5.5ex} UNK users& 16\% & 7\% & 19\% & 25\%\\ \hline
\hline \rule[-2ex]{0pt}{5.5ex} PO posts & 63\% & 79\% & 65\% &  55\%\\ 
\hline \rule[-2ex]{0pt}{5.5ex} PiS posts& 25\% & 17\% & 24\%  &  26\% \\ 
\hline \rule[-2ex]{0pt}{5.5ex} UNK posts& 12\% & 4\% & 11\% & 19\%\\  
\hline 
\end{tabular}
 \caption{Comparison between JAN09, JUL10, FEB11 and FEB11Q user political affiliations. 
  Small differences between ratios calculated for users and for posts reflect generally lower 
  activity of the users whose affiliation is undefined and higher activity of the committed users. 
  This confirms the hypothesis of the forum being used as `battling ground' between the two 
  opposing parties.
  July 2010 data were gathered just after the presidential elections won by PO candidate, when his supporters were elated by the outcome.}
	\label{tab:Affiliations}
\end{table*}

\begin{figure*}[ht]
\includegraphics[width=160mm]{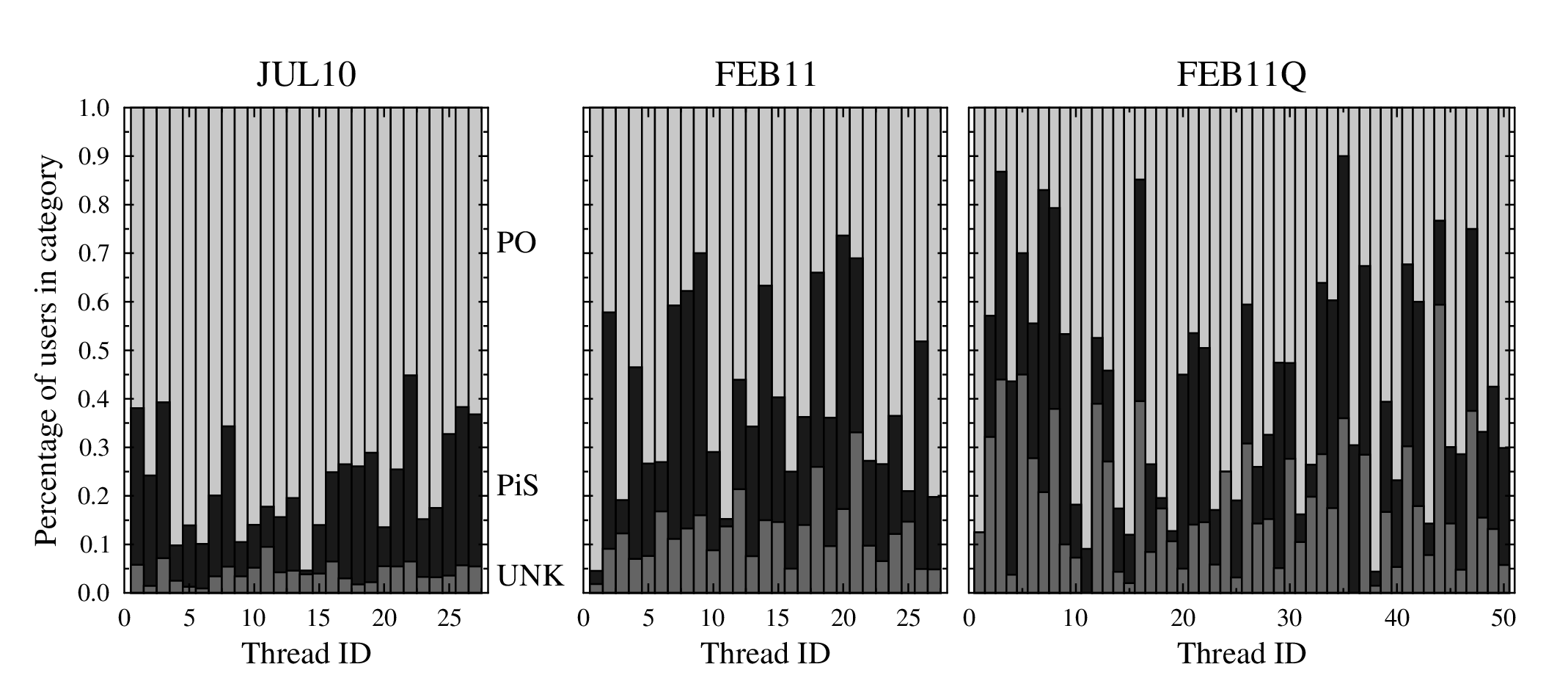}	
	\caption{Distribution of user affiliations for threads in JUL10, FEB11 and FEB11Q datasets. In July 2010 dominance of PO supporters was very high in all studied discussion threads. In February 2011 we observe some threads where there is similar dominance, but also threads where supporters of the two parties are almost in equal numbers or even when PiS supporters dominate. The latter situation is especially interesting, bearing in mind strong pro-PO stance of the newspaper. }
	\label{fig:threadaffiliation}
\end{figure*}

Figure~\ref{fig:threadaffiliation} presents distribution of sympathies of users for each of the 
threads in JUL10, FEB11 and FEB11Q. In July 2010, the polarization of the forum was at the highest 
level in the studied periods, but in all cases there are quite large deviations from the averages 
in particular threads. Despite overall PO dominance, in both datasets from February 2011 there are
threads where PiS supporters posted a majority of comments. 

The UNK category is actually comprised of two distinct groups. The first are the users who openly
declare support for one of the remaining parties in Poland or openly against both PO and PiS. 
The second, roughly the same in size, 
are those users for whom assignment of political support was impossible to determine
from the content of the posts.

User affiliation was determined first within each dataset, and them compared between them. 
Political affiliation has been remarkably stable for the studied forum. Only fifteen users changed
their sympathies, all of these changes happening between July 2010 and February 2011. 
In one case, the change was from PiS to PO support, in another from UNK to PiS. 
Thirteen PO supporters, in all cases dissatisfied with perceived lack of activity by the PO government, 
changed their affiliation, twelve of them switching to support to newly formed political 
parties, and only one declaring support for PiS. 
Taking into account that the number of changes is only 0.23\% of the total number of participants,
we may conclude that participation in political discussions did not encourage change political opinions
and support in general sense.

\section{Discussions content}

One of our goals was to observe if there are characteristic features of the content of posts that persist/change with the passage of time. As in Paper I, we have divided the posts into several categories:
\begin{description}
\item [{Agr}] - comment agrees with the covered material (either the original
news coverage or the preceding comment in a thread);
\item [{Dis}] - comment disagrees with the covered material;
\item [{Inv}] - comment is a direct invective and personal abuse of the previous
commentator;
\item [{Prv}] - provocation - comment is aimed at causing dissent, often, but not always
only weakly related to the topic of discussion;
\item [{Neu}] - comment is neutral in nature, neither in obvious agreement
or disagreement;
\item [{Jst}] - `just stupid' comment, which is unrelated to the
topic of discussion, but without malicious intent;
\item [{Swi}] - comment signifying a switch in participant's position leading to agreement 
between two previously opposing commentators.
\end{description}

Table~\ref{TabAttribH} summarizes the distribution of various types of comments. Firstly 
we note relatively stable ratios of positive (Agr), negative (Dis, Inv, Prv) and
neutral comments. The average values for all datasets are 17\%/70\%/12.5\%, clearly indicating 
that majority of comments are confrontational in nature. 
There is significantly lower number of posts classified as neutral in JAN09 data. 
We observe an increase of neutral comments as we move to later datasets.
Another observation is that the old interface discussions contain more agreeing comments than the 
new interface sets, a finding which we shall discuss in more depth later on. 

Looking at the distribution of comments linking within and between supporters of both parties, 
we observe
that the largest number of links are between the groups supporting conflicted parties (inter-faction),
rather than within each group (intra-faction). This stands in contrast with observations of 
\citet{adamic05-1,hargittai07-1} who analyzed links between political blogs in the United States.
The main difference is that blog entries are, at least in principle, deliberative in nature. 
The process of creation of the network is also radically different. 
The links between blogs are inserted by the authors as integral part of the blog text; 
to support the presented point of view. 
Moreover, many blog entries combine multiple topics, data sources and links.
In contrast, quick comments in discussion fora are usually focused on one topic 
(either the original source or some previous comment) and are rather reactive.  
The links are usually to a single comment only, with very few exceptions.
It is easier (and more rewarding emotionally), in a short time and space, to attack an opponent than to construct elaborate presentation of one's own position by linking with other supporters.
In fact, in research focused on blog content rather than on simple presence of direct links, 
even this more deliberative medium shows dominance of attack approach. 
As \citet{trammell06-1} noted in the analysis of blogs during the US 2004 elections, 
more than a half of the blogs discussed the opponent, and out of these, 
almost 80\% contained an attack.

\begin{table*}
\begin{tabular}{|l|c|c|c|c|c|c|c|c|c|} \hline
 Connection type               & Agr  & Dis  & Inv  & Prv  & Neu  & Jst  & Swi  & \textbf{Subtotal} & $\alpha$ factor \\ \hline
	           \multicolumn{9}{c}{JAN09} \\ \hline
Intra-faction (PO-PO, PiS-PiS) &16.9\%& 2.1\%& 2.1\%& 1.1\%& 1.1\%& 0.4\%& 0.1\%& 21.9\% & 0.51 \\
Inter-faction (PO-PiS, PiS-PO) & 0.6\%&32.8\%&17.1\%& 2.7\%& 0.8\%& 0.0\%& 0.0\%& 54.0\% & 2.00\\
Factions-UNK (PO/PiS-UNK)      & 2.5\%&11.1\%& 3.1\%& 1.9\%& 1.7\%& 0.4\%& 0.2\%& 20.9\% & 1.52\\
Comments by UNK                & 0.6\%& 1.2\%& 0.7\%& 0.2\%& 0.2\%& 0.1\%& 0.1\%&  3.1\% & 0.19\\ \hline
\textbf{ Subtotal}             &20.6\%&47.2\%&21.1\%& 5.9\%& 3.8\%& 0.9\%& 0.4\%& & \\ \hline
	           \multicolumn{9}{c}{JUL10} \\ \hline
Intra-faction (PO-PO, PiS-PiS) &12.4\%& 1.4\%& 1.0\%& 1.8\%& 5.4\%& 0.3\%& 0.0\%& 22.5\% & 0.36 \\
Inter-faction (PO-PiS, PiS-PO) & 1.1\%&28.1\%&25.8\%& 8.2\%& 8.8\%& 0.0\%& 0.0\%& 72.0\% & 3.07\\
Factions-UNK (PO/PiS-UNK)      & 0.8\%& 0.9\%& 0.8\%& 0.4\%& 0.9\%& 0.0\%& 0.0\%&  3.8\% & 0.59\\
Comments by UNK                & 0.2\%& 0.4\%& 0.1\%& 0.0\%& 0.9\%& 0.0\%& 0.0\%&  1.7\% & 0.19\\ \hline
\textbf{ Subtotal}             &14.7\%&30.9\%&27.7\%&10.5\%&16.0\%& 0.3\%& 0.0\%&  &\\ \hline
	           \multicolumn{9}{c}{FEB11} \\ \hline
Intra-faction (PO-PO, PiS-PiS) & 9.8\%& 1.7\%& 1.7\%& 2.1\%& 6.4\%& 0.0\%& 0.0\%& 21.8\% & 0.52\\
Inter-faction (PO-PiS, PiS-PO) & 0.8\%&26.1\%&18.8\%& 7.3\%& 6.5\%& 0.0\%& 0.0\%& 59.6\% & 2.44\\
Factions-UNK (PO/PiS-UNK)      & 1.1\%& 3.5\%& 1.6\%& 1.1\%& 1.5\%& 0.0\%& 0.0\%&  8.9\% & 0.59\\
Comments by UNK                & 0.5\%& 3.8\%& 1.9\%& 0.7\%& 2.7\%& 0.2\%& 0.0\%&  9.8\% & 0.53\\ \hline
\textbf{ Subtotal}             &12.4\%&35.2\%&24.0\%&11.2\%&17.0\%& 0.2\%& 0.0\%&  &\\ \hline
	           \multicolumn{9}{c}{FEB11Q} \\ \hline
Intra-faction (PO-PO, PiS-PiS) &15.6\%& 2.2\%& 0.8\%& 3.8\%& 6.5\%& 0.4\%& 0.0\%& 29.2\% & 0.99\\
Inter-faction (PO-PiS, PiS-PO) & 0.3\%&18.9\%&11.7\%& 3.9\%& 2.4\%& 0.0\%& 0.0\%& 37.2\% & 1.41\\
Factions-UNK (PO/PiS-UNK)      & 2.3\%& 5.7\%& 2.0\%& 1.2\%& 2.6\%& 0.3\%& 0.0\%& 14.0\% & 0.74\\
Comments by UNK                & 3.3\%& 8.4\%& 1.7\%& 1.9\%& 4.1\%& 0.2\%& 0.0\%& 19.6\% & 0.77\\ \hline
\textbf{ Subtotal}             &21.4\%&35.2\%&16.2\%&10.8\%&15.6\%& 0.9\%& 0.0\%&  &\\ \hline
	\end{tabular}
	\caption{Statistics of comment type between various groups of users for the studied datasets (two identified factions  and neutral or unidentifiable class UNK).  Only comments linked to other comments are classified. The $\alpha$ factor denotes the ratio of observed number of comments linking within faction group, between groups and to/from agents with unknown affiliation to values expected from the user affiliations if the posts were placed randomly. For example, for inter-faction comments we observe $\alpha$ value higher than 1, which means that users prefer to address the supporters of the opposite faction. Value smaller than 1 corresponds to lower preference for placing a comment.
	\label{TabAttribH} }
\end{table*}

To analyze this issue further we have calculated the ratio of observed links in each category 
(inter-faction, intra-faction, to and from users of unknown sympathies) to the number of links
expected if one assumes no preferences in commenting. 
Such numbers would be given by the appropriate ratios of PO and PiS supporters in each dataset.
The resulting ratio, named $\alpha$ factor, in presented in the last column in Table~\ref{TabAttribH}.
In all datasets, we observe that intra-faction $\alpha$ is less than 1, which means that there 
is much less motivation to post comments  addressed to one's own group members.
Conversely, the inter-faction $\alpha$ is always greater than 1, reflecting the willingness 
to start/continue discussions with the opponents. We note that in the most heated debate, JUL10, 
which took place just after the presidential elections,
the inter-faction $\alpha$ is greater than 3 and for intra-faction it is equal to 0.36. 
This documents very high disassortativeness shown by the commentators who took the trouble 
to address their posts to other users, despite the lack of the easy tools in the new interface.

Interestingly, the less popular, old interface discussions in  FEB11Q show significant 
difference from the other sets. There is much smaller number of abusive comments (Inv).
Also, $\alpha$ factors are much closer to 1, indicating less preferential linking to political opponents.
As FEB11Q dataset shares a lot of network properties with JAN09, and, at the same time, it is based on
the same newspaper content as the FEB11 dataset, this diminished assortativity is probably due to 
changes in the interface and change in general user attitudes between 2009 and 2011.
We will return to these issues in the conclusions of the paper.

\section{Analysis of emotions}

\subsection{Human analysis of emotions and other users' evaluations}

In Paper I our analysis focused on the content of the user comments, classifying 
them mainly according to the expressed views and opinions as presented in the previous section. 
Here we are expanding the analysis by looking at emotions expressed by user comments. 
Such emotions have been recently the subject of intensive studies (e.g. 
\citet{schweitzer10-1,prabowo09-1,mitrovic10-2,chmiel11-1}), aimed at quantitative description of 
emotional motivations. 
While there is a rough mapping between the goals of the comments (informative, provocative, quarrelsome) and the associated emotions, we note that these categories are not completely the same. Posts categorized as agreements may be expressed in highly emotional fashion -- or stated neutrally. 
Similarly, disagreement may be stated in with or without agitation. Even invectives may be expressed through vulgar, impolite language or through sarcasm, in a cold and calculated  manner. The two directions of analysis: content and emotion are thus complementary.

The method used in this work is an extension of the approach used in \citet{chmiel11-1}. Instead of simple $+1$, 0, $-1$ scale of emotions indicating positive, neutral and negative emotional expressions,  we use a graded  scale from $+1$ to $-3$, described below:
 \begin{description}
\item[+1] Positive emotions, expressed as support for another post or statement by politician described in the news item.
\item[0] Neutral emotions: statements of facts (either agreeing or disagreeing with the target of the post), explanations; worded in neutral language.
\item[-1] Light negative emotions: expressed via single instances of vulgar language, suggestions that politicians or other users are idiots, etc.
\item[-2] Strong negative emotions: repeated use of vulgarities, comparing politicians or other users to Nazi/Stalinist personages, excessive use of capitalized letters etc.
\item[-3] Excessive negative emotions: posts combining many elements described above.
\end{description}
The categorization of emotions in each post is done by human reading. Emotions are assigned to two categories: related to general political issues and politicians (e.g. `I hate politician X') or directed at other users of the forum (e.g. `You are an idiot'). The total emotion expressed by a post is then calculated as sum of the two constituents.

\begin{figure*}[ht]
\includegraphics[width=160mm]{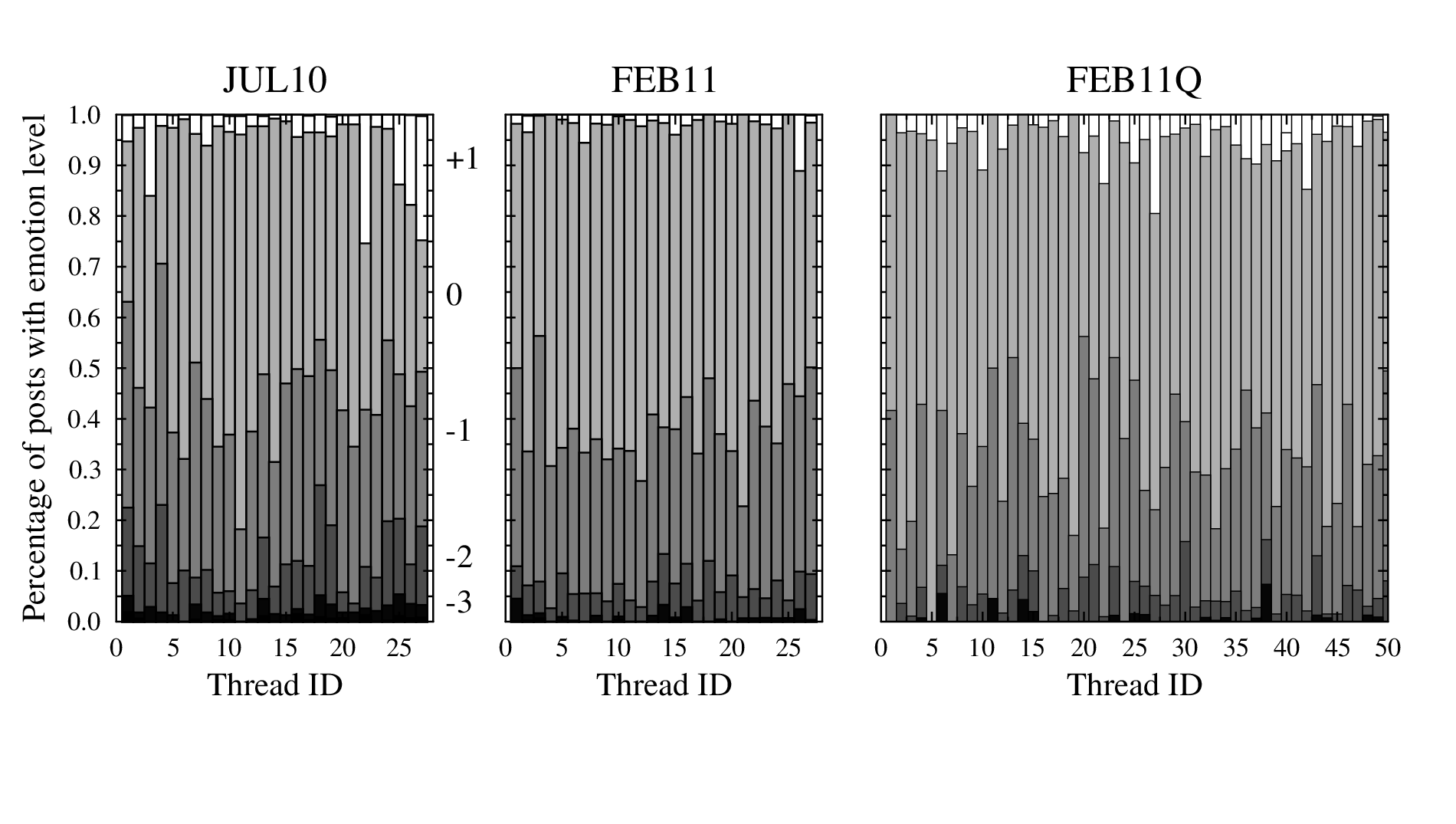}	
	\caption{Distribution of emotions within threads chosen for analysis in JUL10 and FEB11 datasets. Figure shows percentages of posts within each thread with total emotions  ranging from $+1$ to $-4$ (the latter is possible if one post expresses negative emotions against both politicians and forum users). 
Darkening shades of gray represent emotions from  $+1$ to $-4$, indicated also for clarity in the JUL10 dataset.	
}
	\label{fig:threademotiondistrib}
\end{figure*}

Figure~\ref{fig:threademotiondistrib} presents distribution of emotions expressed in posts JUL10, FEB11 
and FRB11Q datasets. It is worth to note much larger ratios of positive comments in threads 3, 22 and 
25--27 in JUL10 set. In all these cases they were concerned with news items regarding comments strongly 
against PiS politicians made by perceived `outsiders' of the political field. In one case (thread 22) 
this was a comment by ex-prime minister from Social-Democrat Left Alliance party. In  four other cases, 
the positive response was directed at an `\textit{enfant terrible}' of PO, 
who some time later decided to form 
his own political party. Positive emotions were mostly of the form of personal support for courage and 
decisiveness of single persons, perceived as acting outside political establishment.
In FEB11 dataset there is only one discussion with high ratio of positive emotions. In thread 26, 
positive emotion has been generated by relatively large number of PiS supporters, expressing the 
admiration for statements by PiS politician accusing  the PO government of treason. 
A reverse situation is present in threads 27 and 32 of FEB11Q set: here the PO supporters express their 
positive emotions at statements by the prime minister and by a popular sportsman. 

\subsection{Automatic emotion recognition}\label{sec:automatic}
In addition to assignment of emotions to comments based on human 
reading, we have constricted simple analysis engine to detect emotions 
from the texts. Engine calculates emotion based of content of the post,
length of the post and average of previous emotion of posts made by user. 

Word list is built using two separate sources. The first contains popular polish swearwords.
The second  is built from words commonly used in community to make fun and/or irritate other users.
Words in the second list are taken from subset of dataset (three randomly selected threads, about
600 posts). 
They include intentional misspellings of names and word-games on political party names, turning them
into near-swearwords.
Each word has its own emotion power property $P_n$, where $P_n$ is integer ranging 
from $1+$ (positive) to $-3$ (very negative), assigned before the evaluation process.

Emotion detection algorithm works as follows:
\begin{itemize}
	\item it reads a comment, divides it into words and counts them, and remembers count of 
	the words as $L$;
	\item it searches for words from the list, each hit adds $0.8P_n$ to post score $S$. 
	We also store total hits count $H$;
	\item average score $U_{i}$ for the author of the post is calculated as arithmetic mean of 
	her/his previous average $U_{{i-1}}$ and current score of the post $S$;
	\item emotion value of the post is calculated from score through 
	$E=S\left(\ln{({25H}/{L} + 2)}\right)^2-U_{i}/5$, 
	this form of expression was chosen to arrive at best reproduction 
	of the results for the three threads used in training of the system;
	\item to allow for situation where a single highly negative word is used in a very long post 
	(which `dilutes' the emotional impact of the post) or when previous posts of the user 
	were overwhelmingly negative, the negative emotion is reduced by looking at ratio of the 
	words from the list found in comment to its length $H_p = H/L$. For $H_p < 0.09$ if negative 
	emotion was lower than $-1$ it  is reset to  $E = -1$, for $H_p < 0.20$ negative emotions 
	lower than $-2$ are reset to $E = -2$.  
	\item all the other results of $E$ are rounded to the nearest integer value.
\end{itemize}

Human and automatic emotion recognition values were compared via average dataset emotion value and through distributions of posts with total emotion equal to $+1$, 0, $-1$, \ldots, presented in  Table~\ref{tab:automatichuman}.
The automatic algorithm, although `trained' only on a very small subset of our data, gives reasonable results 
when compared to human analysis.

\begin{table*}
\begin{tabular}{|p{5.0cm}|p{1.3cm}|p{1.2cm}|p{1.2cm}|p{1.2cm}|p{1.2cm}|p{1.2cm}|}
\hline Dataset & Average emotion & \multicolumn{5}{c|}{Ratio of comments with emotion of}\\
 & & $+1$ & $0$ & $-1$& $-2$& $<-3$  \\ 
\hline JUL10, human analysis & -0.539 & 0.059 & 0.497 & 0.321 & 0.099 & 0.023  \\ 
\hline JUL10, automatic analysis & -0.530 & 0.039 & 0.503 & 0.345 & 0.090 & 0.023  \\ 
\hline JUL10, simulations average & -0.485 & 0.009 & 0.423 & 0.410 & 0.076 & 0.003  \\ \hline
\hline FEB11, human analysis & -0.460 & 0.017 & 0.585 & 0.330 & 0.060 & 0.010  \\ 
\hline FEB11, automatic analysis & -0.490 & 0.026 & 0.557 & 0.312 & 0.088 & 0.017  \\ 
\hline FEB11, simulations average & -0.383 & 0.045 & 0.544 & 0.392 & 0.017 & 0.000  \\ \hline
\hline FEB11Q, human analysis & -0.338 & 0.046 & 0.630 & 0.272 & 0.045 & 0.006  \\ 
\hline FEB11Q, automatic analysis & -0.390 & 0.041 & 0.594 & 0.309 & 0.043 & 0.011  \\ 
\hline FEB11Q, simulations average  & -0.351 & 0.048 & 0.564 & 0.375 & 0.012 & 0.000  \\ 
\hline 
\end{tabular} 

 \caption{Comparison between automatic and human detection of emotions for various datasets. 
The parameters of the automated
analysis algorithm as described in Section~\ref{sec:automatic} have been tuned to give best fit to just three
threads in JUL10 dataset. Despite this, the results for both the average emotion and for the assignment of emotions of posts into different levels of emotion seem quite good. }
	\label{tab:automatichuman}
\end{table*}

The quality of the automatic procedure may also be measured by correlations between 
automatic and human assignment of emotions to all posts in a  forum and for thread averages.
For the JUL10 dataset, correlation coefficients between human and automatic assignment of emotions were calculated for the whole set of posts ($C_{p}^{JUL10}=0.18$)  and for  threads ($C_{T}^{JUL10}=0.60$). 
For the FEB11 dataset the correlation coefficients were ($C_{p}^{FEB11}=0.26$)  and for  threads ($C_{T}^{FEB11Q}=0.54$), while for FEB11Q the respective values are $C_{p}^{FEB11}=0.20$  and $C_{T}^{FEB11Q}=0.57$.

\section{Reader evaluation of posts: tracing the invisible users}

The users of Internet fora are not limited to active authors of comments. There are those who read the messages but do not post comments. Internet slang calls these users `\textit{lurkers}', and usually there is little that we can do to study the statistical properties of this group.

Fortunately, in our case, in addition to \textit{post factum} study of the entire dataset, 
we may use data on how \textbf{forum readers} evaluated the posts shortly after they were written. 
This feature is present in the 
`new interface' datasets (JUL10, FEB11) thanks to thumbs-up/thumbs-down buttons. This provides a 
measure of reactions of readers to the views presented by other users. 
Similar measures are available at other sites, for example as tools in establishing usefulness 
of user advice in self-help fora or in review discussion boards. 
In the context of general discussions reader
evaluations were analyzed  by \citet{gomez08-1}  and \citet{gonzalez-bailon10-1} for \texttt{slashdot}; 
by \citet{hsu09-1,jamali09-1,khabiri09-1} for \texttt{digg}, 
and by \citet{lange07-1} for \texttt{YouTube}.

We used two numbers for each post: 
the total number of evaluations it has received, $N_P$, and resulting 
reader opinion (difference between thumbs-up and thumbs-down votes), $O_P$. 
The first value measures the interest that given post has received from reader community. 
The second is result of an interplay between post political positioning and distribution of political 
sympathies of reader community. Analysis of the correlations between the two variables may provide some 
information regarding the `invisible' part of the forum users: readers who do not post. On average, in 
JUL10 dataset each post has received 10.8 evaluations; in FEB11 the number is smaller, 7.6. 
The fact that there are many more thumbs-up/thumbs-down clicks then posts agrees with an 
expectation that the number of lurkers is greater than the number of writers. 

\begin{figure*}[ht]
\includegraphics[height=140mm]{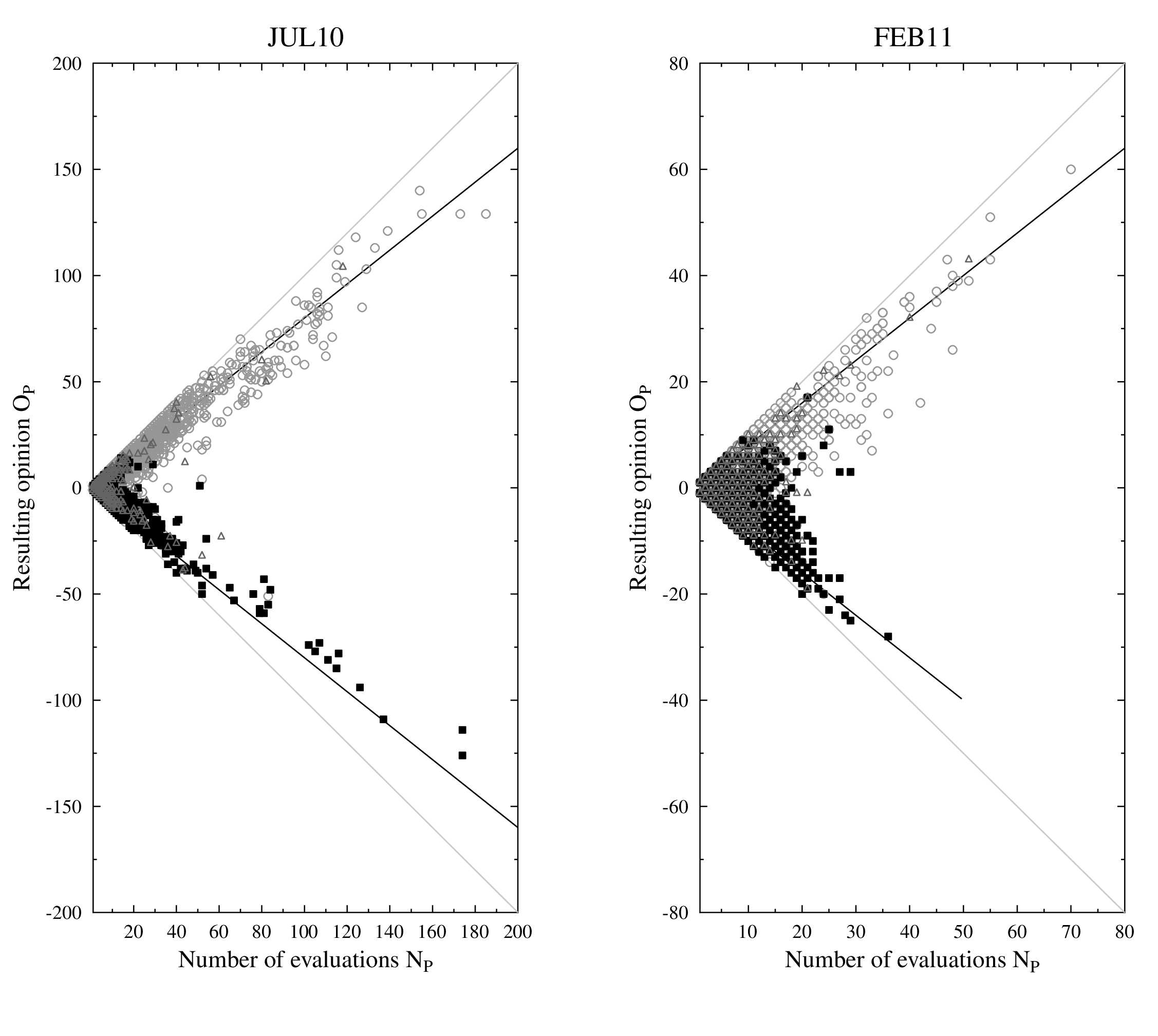}	
	\caption{Evaluations of user posts in JUL10 (left) and FEB11 (right) datasets. 
	All points correspond to single posts. Different markers indicate posts by supporters of political 
	parties (light gray circles majority PO users, black squares minority PiS, triangles -- unknown 
	affiliation). Gray lines show boundaries of fully negative/positive evaluations. Black lines are 
	guide for the eyes, with $O_P =\pm 0.8 N_P$. 
	Assuming that only politically  committed \emph{readers} would give the evaluation, 
	and that PO supporter would always approve pro-PO post and always disapprove pro-PiS post, 
	and vice versa, the lines correspond to 90\% of the evaluating readers being pro-PO, 
	much higher proportion than for the comment writing forum users.   }
	\label{fig:twoevaluations}
\end{figure*}

Plotting the number of evaluations $N_P$ vs. resulting opinion $O_P$ 
(Fig. \ref{fig:twoevaluations}) shows 
interesting behavior. The distribution splits into two separate `wings' of positive and negative 
opinions scaling approximately as linear functions of the number of evaluations. The absolute value of 
coefficients are somewhat below 1. The number of posts in the positive `wing' is greater than in the 
negative one, and, at least for the posts with highest numbers of evaluations, negatively judged ones 
are written by PiS supporters, whereas the positively evaluated ones by pro-PO users. 
For posts which received many evaluations, one can approximate $O_P\approx \pm 0.8  N_P$  for PO and 
PiS supporting posts respectively (indicated as black lines in Fig.~\ref{fig:twoevaluations})
Using a simple assumption that evaluations are 
done by the supporters of the two conflicted camps,  and that supporter of the 
post political view would always give positive opinion 
and vice versa, the PO supporter \textit{reader} ratio would be at 90\%, higher than the 
observed \textit{writer} ratio.

\begin{figure*}[ht]
\includegraphics[width=160mm]{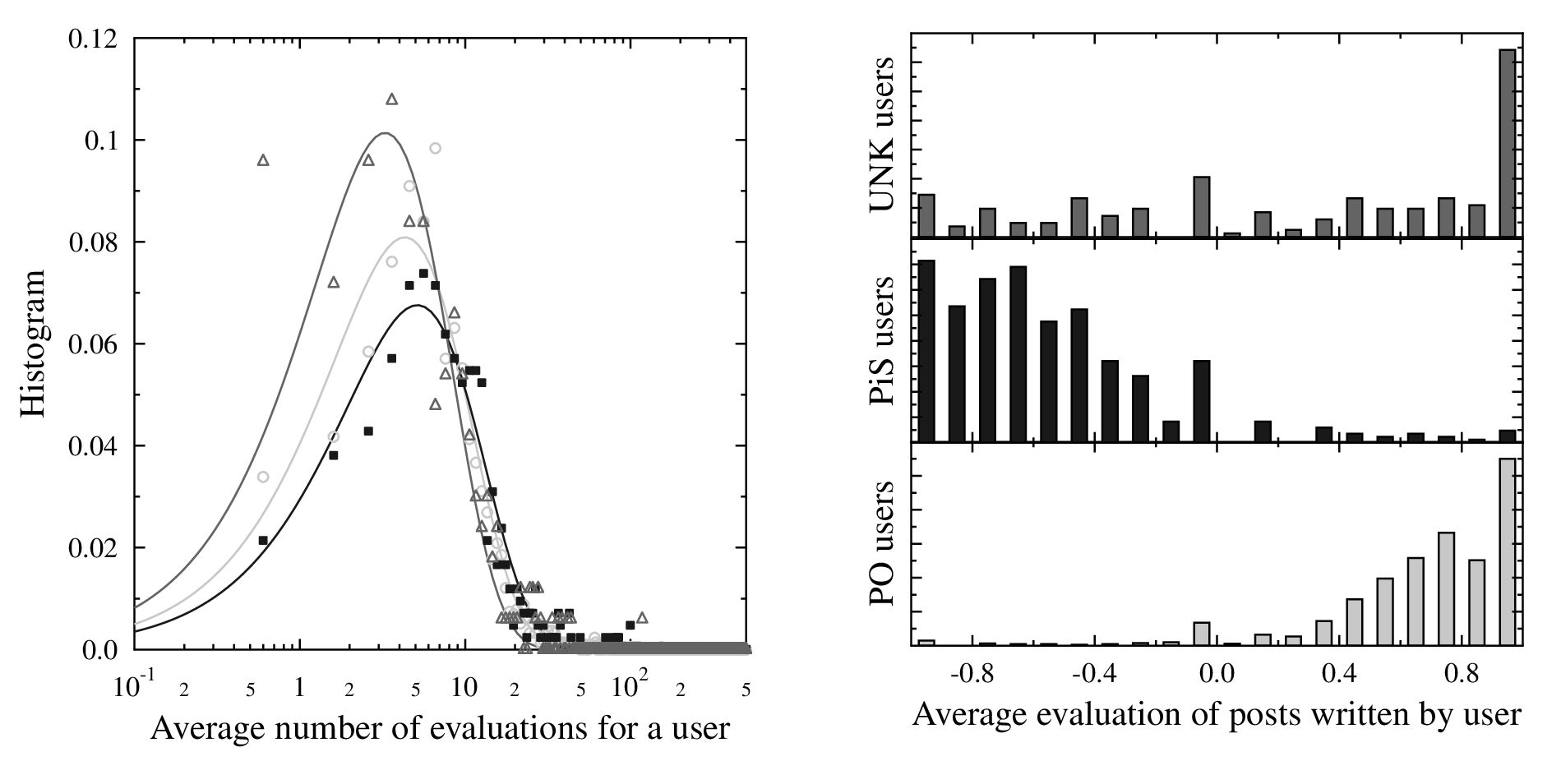}	
	\caption{Statistics of user evaluations as function of their affiliation in JUL10 dataset. Left panel: normalized histograms of number of evaluations received by a user, for the three classes os users (PO supporters: light gray circles, PiS supporters black squares, unknown triangles). Lines are fits with $ax\exp(-\lambda x)$, similar results obtained for each user group. Right panel: histogram of average evaluation of posts written by a user, the same color scheme applies.  }
	\label{fig:JULuserevaluations}
\end{figure*}

\begin{figure*}[ht]
\includegraphics[width=160mm]{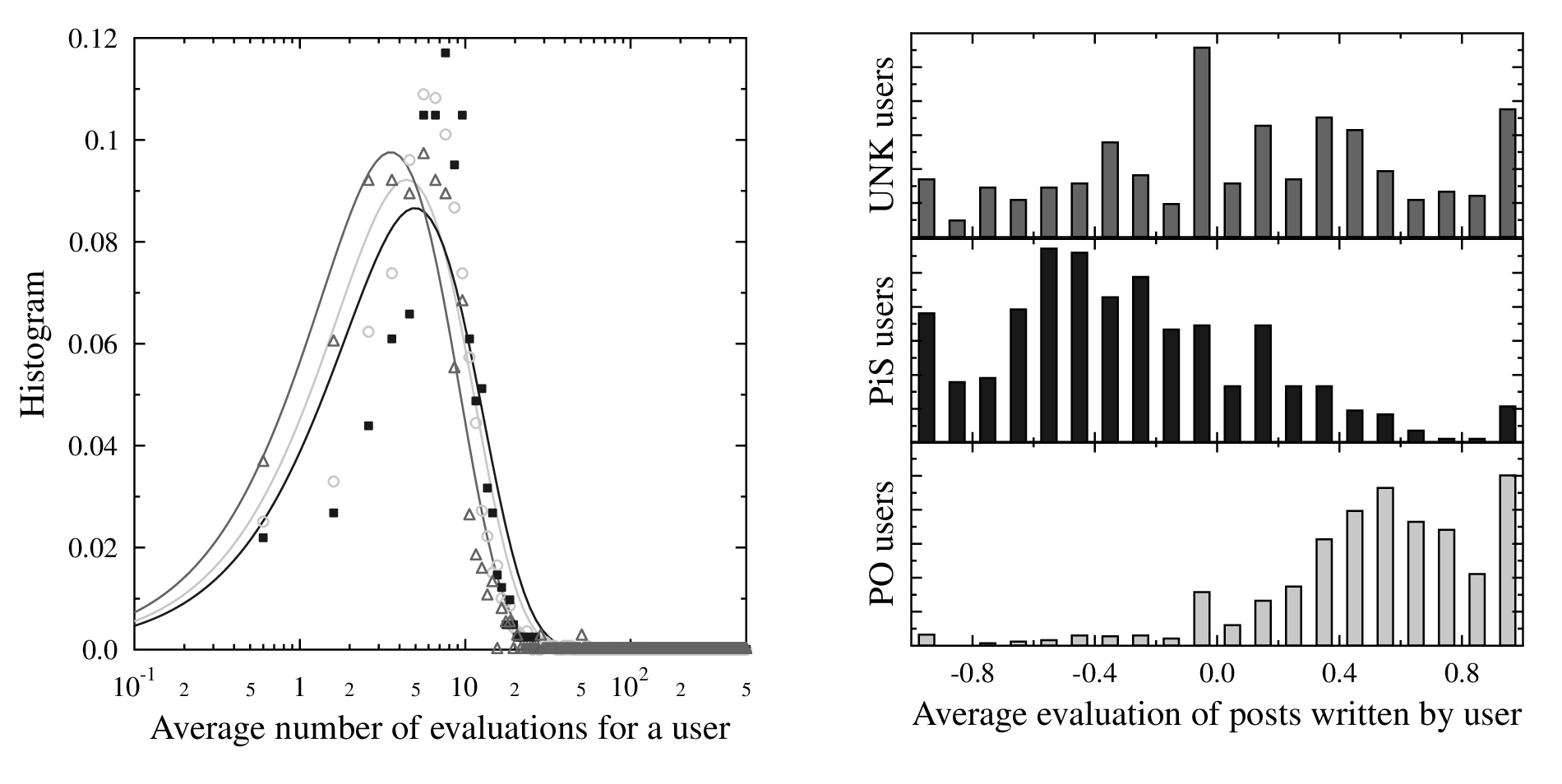}	
	\caption{Statistics of user evaluations as function of their affiliation in FEB11 dataset. Left panel: normalized histograms of number of evaluations received by a user, for the three classes os users (PO supporters: light gray circles, PiS supporters black squares, unknown affiliation -- triangles). Lines are fits with $ax\exp(-\lambda x)$, similar results obtained for each user group. Right panel: histogram of average evaluation of posts written by a user, the same color scheme applies. }
	\label{fig:FEBuserevaluations}
\end{figure*}

We have also checked whether the number of evaluations received depends on the post/user political 
views. Left panels of figures~\ref{fig:JULuserevaluations} and \ref{fig:FEBuserevaluations}  present 
normalized histograms of the average number of evaluations per post for users supporting PO, PiS 
and of unknown sympathies. In each case these distributions are well described by 
$ax\exp(-\lambda x)$ functions, similar for each political group. 
On the other hand, the histogram of opinions (right panels of the figures ~\ref{fig:JULuserevaluations} 
and \ref{fig:FEBuserevaluations}) shows radical differences between supporters of different parties. PiS 
supporters, who are minority in the forum receive very few positive opinions and distribution increases 
towards totally negative $-1$ opinion/number of evaluations ratio. Opposite phenomenon is seen for PO 
supporters. Unknown users are as likely to get positive as negative opinion.

\section{Agent based modeling of the user activities}
In Paper I we have presented a computer model of participation in political discussion fora. 
The model was further developed and applied to BBC discussion fora in \citet{chmiel11-1}. 
We have used the same model framework for the four datasets studied here. 
Thanks to full scope human analysis of the posts some of the parameters for the model, 
which were guessed or fitted in previous studies could be input from observations or used to compare the 
simulations to real world, for example the ratio of proponents of the 
two competing parties and of the neutrals/unknown 
affiliation users.
The model uses, as input parameters, probabilities of posting a comment, 
addressing it to source or to other post,
probability of response within a pairwise exchange between users 
(depending on whether it is written by a user sharing political sympathies or not).
Because simulations use probabilistic distributions and give  slightly different results at each run,
we present results as averages of 20 such runs for each dataset. 
Table~\ref{tab:simulations}  presents average number of users, threads above length threshold, posts 
and links, as well as the distributions of connections between different user groups.

\begin{table*}[ht]
\begin{tabular}{|p{6cm}|p{2cm}|p{2cm}|p{2cm}|p{2cm}|}
\hline \rule[-2ex]{0pt}{5.5ex}   & JAN09 & JUL10 & FEB11 & FEB11Q\\  \hline
\hline \rule[-2ex]{0pt}{5.5ex} Number of threads & 45 & 28 & 30 &  55\\ 
\hline \rule[-2ex]{0pt}{5.5ex} Number of posts & 7315 & 7073 & 6485  &  4560\\ 
\hline \rule[-2ex]{0pt}{5.5ex} Number of users & 1557 & 2691 & 2168 & 1627\\ 
\hline \rule[-2ex]{0pt}{5.5ex} Number of links between users (as percentage of posts) & 4576 (62.5\%) & 748 (10.5\%) & 1148 (17.7\%) & 2299 (50.4\%)\\ \hline
\hline \rule[-2ex]{0pt}{5.5ex} Intra faction links percentage & 22\% & 23\%& 27\%& 25\%\\ 
\hline \rule[-2ex]{0pt}{5.5ex} Inter faction links percentage & 53\% & 68\%& 45\%& 37\%\\
\hline \rule[-2ex]{0pt}{5.5ex} Factions to UNK links percentage & 12\% &  5\%& 13\%& 17\%\\
\hline \rule[-2ex]{0pt}{5.5ex} Comments by UNK percentage & 13\% &  4\%& 15\%& 21\%\\
\hline 
\end{tabular}
 \caption{Main results of simulations of discussions (averages over 20 simulations of each dataset). 
 For numbers of threads, users, posts and links, compare with Table 1, for the distribution of connections between
 supporters of political parties, compare with column Subtotal in Table 3.}
	\label{tab:simulations}
\end{table*}

The network properties (in-degree, distribution of the number of written posts 
and number of threads a user participates in) of simulated discussions  are 
compared with observations in figures~\ref{fig:indegree4}--\ref{fig:userthreads}. 
With the exception of the distribution of the number of threads a user participates 
in for the FEB11 set, simulations reasonably reproduce multiple system characteristics, 
such as observed distributions reasonably well.

We have extended and changed the model used in \citet{chmiel11-1} with respect to the evaluation of emotions. 
Previous approach assumed a fixed reaction of an author in response to political 
stance of the author of commented post: the resulting emotion was $+1$ if their 
affiliations were the same, $-1$ if they belonged to opposite camps and either $+1$ or $0$ 
if a committed agent commented a message written by a neutral (neutral agent's 
comments were assumed to be always neutral).

Thanks to a deeper human analysis of our data we were able to propose a modification of this approach. 
The reaction of an agent to a post is in simulations is taken as the nearest integer from Gaussian 
distribution with center and width determined by comparing the political views of the two agents. 
This model is based on observed the statistics of posts, where data for various author-target 
affiliation combinations were fitted to normal distributions. As it turned out, the distributions 
could be divided into three groups,
differing by position of the center and width of the distribution.
The first corresponds to similarity of views: PO-PO, PiS-PiS, 
PO-Source (we assume all Source messages are pro-PO). The second is when the author and 
target "disagree" (PO-PiS, PiS-PO, PO-UNK, PiS-UNK). The third, "neutral" distribution was 
observed for posts written by authors with no identified affiliation, regardless of the target.

For the purposes of the simulations, small differences within each group were looked over and post emotion was 
drawn from normal distribution generalized for the appropriate  author-target class. 
The values were those taken 
from observations for each of the datasets separately. 
For example, emotion in similarity case for JUL10 data was 
centered at $-0.4$, for disagreement at $-0.8$ and for comments by unknown agents at $-0.3$; in all 
cases the width of distributions was assumed, $\sigma=0.7$. Due to rapid decrease of the normal distribution, the 
resulting distributions do not agree very well for extremely negative comments (with total emotions of 
$-3$ and $-4$) but give rather good values for the less emotional posts statistics.

\section{Discussion and conclusions}

Looking at temporal changes in the observed discussion properties we note that the 
JUL10 data has the highest negative emotion content. 
As noted before, this dataset contains discussions which took place when the entire public debate was 
at its hottest, shortly after PiS candidate lost his presidential bid. During the election campaign, 
the candidate assumed quiet style, suitable for mourning and reconciliatory stance. Both himself and his 
campaign team refrained from outright attacks on the opponent. After the lost election this 
stance changed radically, and politicians from PiS started immediate, vehement attacks on the 
president-elect and on PO government, accusing them of causing the plane crash 
or suggesting assassination. 
The tone of political debate has been immediately picked up by PiS supporters, 
including the forum users.
At the same time, PO supporters were clearly triumphant after the win. 
This resulted in PO dominance in the dataset.  
Moreover, during the same period the polarization of the whole society led to increased separation of 
`virtual' agoras for the supporters 
of PO and PiS. To check this we have briefly analyzed July 2010 discussions related to 
a well known pro-PiS blog site (\url{kataryna.blox.pl}). In these discussions PiS 
supporters are majority, and we found them to be characterized by mirror image network and emotion statistics.

Without communication the two communities of supporters of PO and PiS 
diverged in worldview, basing their opinions of 
totally separate sets of `factual' data and authorities. 
The process is observed especially in the context of the Smolensk plane crash, 
where the pro-PiS readers have 
gathered around several Web pages containing   claims supporting various conspiracy theories. 
These Web sites bear no relation to the official investigation and are 
cross-linked extensively among themselves, much like 9/11 conspiracy sites in the US. 
As a result the reader focused on such subset of sources 
may be under impression that it presents a coherent and complete view, 
because there is no reference to contrary data and explanations. 
Armed with  one-sided information, the readers are strengthened in their political opinions and less
ready to participate in discussions with opponents, and even less to work for a consensus.

Such behavior has been studied in laboratory environment by \citet{cohen03-2}. 
He has noted that groups not only affect attitudes on key subject but may 
`define the very meaning of the objects in real world'. Opinions of trusted in-group members
are accepted without deliberation, regardless of the merit of the content.

Discussing polarization and fragmentation of society, \citet{stroud10-1} 
has noted that such selective exposure
may have two effects on the population. First, using only the information provided by one side 
would limit a persons ability to deliberate and choose. Second, such separation might lead to 
lesser tolerance and extremism. The existence of Internet discussion fora where conflicted sides are, 
at least, trying to communicate may be a way to combat such extremism.
Allowing the users to interact with each other, even if this interaction is, in many cases, in the form of abuse and invectives may decrease some of the negative emotions. 
This supports the hypothesis put forward by 
\citet{mutz01-1} that media would surpass face-to-face communications across political divides.
The discussions provide much needed exposure to dissimilar views and they complement the 
traditional media in this eye-opening role.
We must, however, remember that simple exposure to the opposite views is no
recipe for open-mindedness. Experimental study by \citet{lord79-1} has clearly shown that 
opinions held before being exposed to other viewpoints strongly bias the reception. 
This bias covers not only debatable opinions but also acceptance of hard, empirical facts.
People would choose only the evidence that supports their views. 
Thus, to achieve effective communication between conflicted groups, 
more is needed than just combining opposing messages in one place.

One of the main observations in this study is that the change of the user interface has led to drastically 
diminished number of inter-user exchanges. As most of these (in the old forum) have been in the form of 
quarrels between supporters of opposing political views, one may be tempted to claim that the interface change 
has led to lessening of direct confrontations and, presumably, level of conflict. This hypothesis is not 
confirmed by comparison of February 2011 data on discussions in two parallel fora with different interfaces.
In fact, the `quarrel promoting' old interface FEB11Q dataset shows much less emotion and more even
distribution of links, indicated lesser recognition of who is it worth communicating with.

There may be several reasons for the less aggressive characteristics of FEB11Q discussions.
The first is smaller role of the \textit{trolls}. These are usually defined as users who post 
many comments aimed at provoking fights. Typical advice for discussion board users is 
`do not feed the troll', and if followed by most participants this leaves trolls with significantly
smaller indegree than the number of posts. There is no formal numerical threshold for `trollness', but 
we may define as a troll any user who posted more than 15 comments in the analyzed discussion set, with 
average emotion less than $-0.3$ and indegree smaller than 1/2 of the posted comments.
Under such conditions FEB11 dataset contained 23 trolls, and FEB11Q only seven. If the threshold is 
more stringent, average emotion less than $-0.4$ and indegree smaller than 1/3 of posts then FEB11 set includes 13 trolls, while FEB11Q none. 

The second difference comes from the opposite end of the user spectrum: one time writers, who 
direct their posts at the source news message. The average emotion of such users in FEB11 set is 
$-0.45$, while for the FEB11Q set is is only $-0.35$. One of the sources of the difference may be the 
anticipation of the scoring mechanism, present in FEB11 forum. Getting a large score is a goal
for many users -- this is clear from the texts of the comments. It is less important whether
the score is positive or negative, what counts is being noticed. This quite naturally motivates
the users to more extreme positions in the forum where the evaluation mechanisms are present.
The reward of watching one's own  comment get noticed and evaluated is complemented by the 
possibility of making  negatively evaluated post literally vanish from the forum, if the number of negative evaluations surpasses certain threshold. 

The two fora differ also in the ways they promote reactions to agreeable/disagreeable post. 
The new interface makes it easy to use the evaluation buttons. Pushing them does not really require 
thorough reading of the post, much less formulating any response. All it requires is a recognition of 
`a friend' or `an enemy'. In contrast, the old interface promotes exchanges of posts. Even though many
of these consist of invectives, many users are forced to actually state what is it that they disagree with. 
Sometimes even explaining in detail why they disagree. Quite often flame and abuse on one side are met with 
calm and deliberation on the other side. While we have not observed \textit{bona fide} opinion changes 
(comments of the Swi category), there were numerous occasions where discussion evolved from
accusations and invectives to explanations and arguments. This would not be possible if the only tool 
at the disposal of the reader is a simple approve/disapprove button.

\citet{slater07-1} has postulated existence of a spiral of selective media use, leading to polarization of beliefs, actions and attitudes, which again influences media use to become more selective.
This process is clearly visible in Polish society, moreover, it has reached another level of positive feedback. 
The whole landscape of the media has changed, adjusting to polarization of the population.
There is a clear division of pro-PO and pro-PiS newspapers, weeklies and TV broadcasters.
The polarization of media is so extreme that one of nation-wide TV stations has never invited any PO
politician (even during four years PO has been in government) while prominently featuring PiS activists.
The same split applies to journalists: lacking any pretense of neutrality and impartiality they have 
grouped themselves in journals that are clearly fighting on both sides of the political barricade. 
The positive feedback driving this change is not only via personal beliefs of the journalists. 
Polarized media sells very well: a new weekly publishing only anti-PO texts has catapulted to second place 
in circulation numbers just after 3 weeks after introduction.

The descriptions of reality that are offered by so polarized media and selected by the users are 
impossible to combine into one coherent worldview. One side treats the plane crash as tragic accident, the other as a terrorist attack in which the government played an active role. For one side Poland is economically healthy and fast growing country, for the other it is on the verge of economical catastrophe or even already collapsed. Such divisions are so deep that they make communication between supporters of the two camps --
especially face-to-face -- extremely difficult, separating the society.
It is interesting to note that in-group influence goes beyond simple conformism,
 but leads also to deep changes in memory, 
as shown by \citet{maede02-1,loftus05-1,wright09-1,edelson11-1}. 
Such effects would influence
the future opinions of the people who see only uniform or mostly uniform opinions
within a political group. 
Thus, the effect of group pressure would have long-term effects beyond those 
characterized by \citet{cohen03-2}. 
Promoting  exposure to contrary opinions, even if it does not lead
to true value judgments  and opinion/attitude change, 
decreases chances of such socially induced distortion of perception and memory, 
simply by observing that there are people with different views.

Participants of the discussion fora are, in this environment, rather special: 
they want to express their views in a way that reaches the other side.
More: they want to communicate to the other side. 
We have already noted that the change of the user interface, making direct responses difficult, 
has been heavily criticized by many users 
from all political camps. In many cases, the users themselves emphatically noted 
that the negative reaction to the change in the user interface 
is `\textit{the only topic on which they agree}', which is confirmed largely by our analysis. 
More than half of the inter-group Agr type comments in JUL10 dataset (shortly after the 
new forum interface has been introduced) are agreements on this particular topic, brought 
spontaneously into otherwise political discussions.
Forum administrators seem to become a common enemy for both political camps, 
because they deny the right to open (though bloodless) fight. 
This may be compared to behavior of football hooligans, 
who fight each other, but unite when confronted by police trying to separate them.
But perhaps there is more positive view of this unity -- the need to talk.

The new forum interface with its diminished capacity for pairwise exchanges is dominated by
`one shot' comments, which are not addressed to a concrete user but rather express 
writer's view to world at large. 
To gain attention these comments are usually more provocative and do not have to refer 
to opposite views. In contrast, when a user posts non-invective reply to another post,
he or she has, at least, to read it. Experimental study of \citet{palmer10-1}
has shown that while simple exposure to negative argumentation leads to more extreme positions, 
this effect may be moderated by actual evaluation of messages. 

In a large scale meta analysis \citet{pettigrew08-1} have identified three ways of 
reducing inter-group prejudice: increasing the knowledge about the other group, 
reducing anxiety related to contact and increasing empathy. 
The study shows that increased knowledge has the weakest effect in diminishing prejudice. 
On the other hand, Internet fora are not good environments to promote empathy:
anonymity of the contacts prevents it.
This leaves anxiety reduction resulting from intergroup contacts as a possible
way of improving communication across political divisions.
In discussion fora such contacts correspond to pairwise exchanges of posts, when users 
`talk' to each other, even hidden behind the anonymity of the forum. 
A reply to a post is treated as personal communication, not as general statement. 

\citet{mackuen10-1} introduce an interesting concept of two idealized types of participants in
political debates: deliberative citizen, who considers all arguments, including these opposite to his views and passionate supporter of one view, the partisan combatant. In real situations
people behave somewhere between these two extremes. \citet{mackuen10-1} argue that it is emotions that define the deliberative vs. combative stance. Moreover, effects of emotions would differ if the source is aversion due to negative feelings such as fear of the other group or 
anxiety, due to lack of knowledge or new situation. Aversion strengthens reliance
on the views already held, while anxiety encourages seeking out more information.
Experimental study has shown that when the situation is seen as familiar threats (leading to aversion)
possibility of a compromise decreases dramatically.
When the situation is treated as novel and challenging people are more open-minded 
and allow compromise to form.

With this in mind we might look at the discussion fora with more hope than fear. 
Despite the generally negative emotions, the discussions provide a way of exposing oneself to
the views of others, and also require participants to formulate their positions in communicable way.
Especially when two users engage in a discussion using explanations rather than invectives. 
Based on the results we suggest that to achieve better results 
in promoting understanding and bridging the 
polarization gap the technology used should facilitate user-to-user exchanges as much as possible.
Instead of anonymous like/dislike polls, the interface should 
emphasize the dialog and also visibility of
such dialogs to other users. It is important that 
onlookers would see that it is possible to communicate
across political division. Of course, large part of these messages would be, 
as we have shown provocative or abusive. But the comparison of FEB11 and FEB11Q 
data shows that bigger fraction of person-to-person 
exchanges diminishes the negative emotional and confrontational status. 
Properly designed Internet discussion fora might become one of the forces 
that pull people together, despite their differences, towards democratically 
desirable goal of participation and communication (\citet{stroud10-1}), even if they
remain fixed in their opinions.
Taking into account observations of \citet{pettigrew08-1} and \citet{mackuen10-1} one might 
postulate a change in the role of moderators. Instead of simply hiding or deleting
posts that are considered to break the rules of communication,
the moderators might visually promote discussions that are based on novel argumentation 
and informative exchanges rather than those using invectives and provocations. 
This would diminish the visibility of trolls and provocateurs, and `set the tone' 
of the forum, without actually banning the extremist views. 
As we argue here, visual presentation plays important role in shaping the reactions of the readers.

Another argument for facilitation of user-to-user comments has been proposed by \citet{kelly09-1}.
He argues that extremist users who have nothing interesting to 
state are actually avoided by other users,
in a form of self-moderating mechanism. 
Such users, called by Kelly `fringe', showing some of the troll-like characteristics, 
were identified by specific
network linking patterns by analyzing \texttt{Usenet} discussions. 
The analysis of their contributions has shown that they were, indeed, 
social undesirables, whose contributions consisted of hatred and racism.

Comparing Kelly's observations with our data, we did not observe `fringe' users, their role taken by 
trolls (who, however, has much greater acceptance level within their own camp). Importantly, only 
fora with easy user-to-user communication allow the user community to isolate undesirables. 
Without such communications all users are left with the option of writing to general audience, 
without the feedback of responses.
And to get the responses, posts have to be interesting (even if abusive or provocative) which 
promotes deliberation. This is exactly the optimistic conclusion of \citet{kelly09-1}.

Lastly, we must comment on a particular weakness of our work that has surfaced recently.
Throughout the study we have assumed that all users participated in the discussions because they
wanted to freely express their views and reactions to other users. In May 2011, one of newspapers
has announced\footnote{Polska The Times, May 30th, 2011.} that 
Polish state prosecutor is conducting an investigation 
covering the use  of paid services, 
consisting of posting inflammatory comments on discussion boards,
by political parties and associated institutions. The phenomenon, 
was reported to cover many Web sites, and has supposedly become `a regular industry', 
with rates of 0.1-0.5 USD per posted comment. 
Indeed, posts containing accusations of being `on the payroll' of PO or PiS are quite 
frequent in the studied datasets, either aimed at particular users on in general,
but there were no proofs. 
Should the investigation prove that the phenomenon is truly widespread, 
we would face the situation where some of the conclusions (although not the raw observations) 
presented in this paper would have to be reconsidered. 
At least part of the emotions expressed by the comments would be faked, scripted 
according to some Public Relations instructions prepared by political parties. 
Such comments  would rightly deserve being in the `provocative' category.
Some of the users we have treated as separate (and included in statistical measures) 
might turn out to be multiple avatars 
of a smaller number of `professional' participants. 

In such a case our observations would have to be reinterpreted.
The paradigm of free, 
even anarchic, exchange of views between individual members of society, 
would have to be replaced by a game between the `professionals' (representing PR branches 
of the political parties) and the rest of society interested in politics. 
While interesting from research point of view, this picture is a very disquieting one.

%\section*{References}

%\bibliographystyle{model2-names}
%\bibliography{../Bibliography/JetBibliography}

\end{document}